\documentclass[11pt]{article}

\usepackage{times}
\usepackage{geometry}
\usepackage{authblk}
\usepackage{amsmath}
\usepackage{graphicx}
\usepackage{subfig}
\graphicspath{ {./images/} }

\usepackage{tikz}
\usetikzlibrary{shapes,arrows}

\title{\huge \textbf{Predicting Semen Motility using three-dimensional Convolutional Neural Networks}}

\author[1]{Priyansi}
\author[2]{Biswaroop Bhattacharjee}
\author[3]{Junaid H. Rahim}
\affil[1,2,3]{\texttt{\{1905110, 1905832, 1905831\}@kiit.ac.in}}
\affil[1,2,3]{School of Computer Engineering, KIIT}

\date{}

\begin{document}

\maketitle

\begin{abstract}

Manual and computer aided methods to perform semen analysis are time-consuming, requires extensive training and prone to human error. The use of classical machine learning and deep learning based methods using videos to perform semen analysis have yielded good results. The state-of-the-art method uses regular convolutional neural networks to perform quality assessments on a video of the provided sample. In this paper we propose an improved deep learning based approach using three-dimensional convolutional neural networks to predict sperm motility from microscopic videos of the semen sample. We make use of the VISEM dataset that consists of video and tabular data of semen samples collected from 85 participants. We were able to achieve good results from significantly less data points. Our models indicate that deep learning based automatic semen analysis may become a valuable and effective tool in fertility and IVF labs.
	
\end{abstract}

{\bf Index Terms - } Semen Motility Prediction, Residual Networks, 3D Convolutional Neural Networks \footnote{Published in \textit{Project Innovations in Distributed Computing and Internet Technology, 17th ICDCIT.}}

\section{\normalfont \textsc{Introduction}}

The In Vitro Fertilization (IVF) industry in India has been growing annually at a rate of 14.7\%, it is valued at 557 million USD and is projected to be valued at 1453 million USD by 2026. We can credit this growth to our increased sedentary lifestyle, reduced physical activity, high stress etc. Semen analysis is an integral part of IVF, it includes analysing the male semen to check various factors recommended by the WHO like morphology, motility, count, volume, head-tail-mid defects etc. These factors help doctors understand semen quality and recommend further diagnosis to couples. \\

Computer Aided Semen Analysis (CASA) is done using expensive heavy equipment that test the physical sample to get results. In labs that cannot afford these machines, the sample is checked manually under a microscope. Current CASA methods do give fairly accurate results but there are problems with this approach. The results depend on the physical consistency of the sample which is subject to change with time, if the sample is not stored properly then the results vary, there is a chance of mishandling in the lab. Also the time go get the results is relatively high, it usually takes five to six minutes to generate a report, if we consider the time from preparing the sample to report then its around ten to twelve minutes. There has been need for a better methodology. \\

Our project is about using a deep learning based approach to solve this problem and present a rapid and reproducible methodology for semen analysis. We trained spatial-temporal 3D Convolutional Neural Networks to predict sperm motility from a microscopic video of the semen sample. We made use of the VISEM \cite{Haugen:2019:3304109.3325814} dataset, it is made up with videos of semen samples collected from 85 participants, along with that there is tabular data on recorded serum levels of the sample, participant data like BMI, age etc. We trained three Residual Neural Networks \cite{DBLP:journals/corr/HeZRS15} with 18 and 34 layers to take videos and tabular data as input and predict whether the sample is progressively motile, non-progressively motile or immotile. We were successful in obtaining high accuracy with significantly less data points. The best model consistently gave an accuracy of 0.875 with an average inference time of 400 milliseconds.

\section{\normalfont \textsc{Technology Used}}

We used the PyTorch\footnote{https://pytorch.org/} library to design and train the neural networks. It is a python based library used to specify various layers of tensor operations and is a standard tool used in deep learning research and production. We also used other libraries in the python data science ecosystem like Pandas, Numpy\cite{harris2020array}, Matplotlib etc. to perform exploratory analysis on the dataset.

\subsection{\normalfont \textsc{Convolutional Neural Networks}}

A Convolutional neural network (CNNs)\cite{lecun1995convolutional} is a specialized type of network used for processing data that has a grid-like topology. They are most commonly used on image and video data. The core idea of CNNs lies in the convolution operation which operates on two functions, it is usually denoted by the $*$ sign. Analytical convolutions are mathematically defined as follows,

$$s(t) = (x * w)(t) = \int_{-\infty}^{\infty} x(a)w(t-a) da $$

In deep learning, as we are dealing with multidimensional discrete data, we use a discrete version of the convolution operation, for an image represented as a matrix $I$ and a kernel matrix as $K$, it can be defined as follows

$$S(i,j) = (I*K)(i,j) = \sum_m \sum_n I(i+m, j+n)K(m,n)$$

A simple example of the convolution operation would look like

$$
\begin{bmatrix}
	a & b & c & d\\ 
	e & f & g & h\\ 
	i & j & k & t
\end{bmatrix} * 
\begin{bmatrix}
	w & x\\ 
	y & z
\end{bmatrix} = 
\begin{bmatrix}
aw+bx+ey+fz & bw+cx+fy+gz & cw+dx+gy+hz\\ 
ew+fx+iy+jz & fw+gx+jy+kz & gw+hx+ky+lz
\end{bmatrix}
$$

A typical convolutional layer consists of three parts, the convolution operation, a non linear activation function, and finally a pooling function to modify the layer further. One of the most common pooling functions used in CNNs is Max Pooling, its returns the maximum value within the rectangular filter matrix. The overall architecture of a regular CNN is shown in Fig. 1.

\begin{center}
	\begin{figure}
		\includegraphics[width=0.95\textwidth]{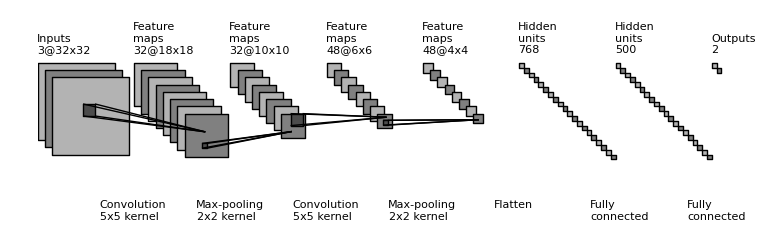}	
		\caption{Architecture of a regular Convolutional Neural Network with 5 convolutional layers}
	\end{figure}
\end{center}

\subsection{\normalfont \textsc{Residual Neural Networks}}

Deep learning has greatly benefited from deeper models. But one of the problems with training deeper models is vanishing/exploding gradients during backpropagation. Residual neural networks (ResNets) \cite{DBLP:journals/corr/HeZRS15} helps us train deeper networks by addressing this degradation problem. Instead of fitting a few layers to a desired underlying mapping, ResNets explicitly lets these layers fit a residual mapping. The core idea of ResNet is introducing a so-called “identity shortcut connection” that skips one or more layers, a diagrammatic representation is shown in Fig. 2.

\begin{figure}
	\centering
	\includegraphics[width=0.5\textwidth]{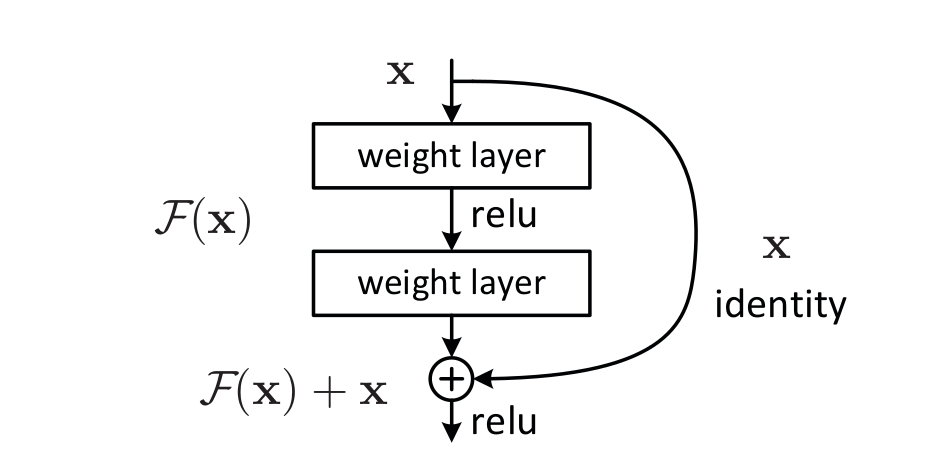}	
	\caption{Building block of a Residual Neural Network}
\end{figure}

\section{\normalfont \textsc{Literature Review}}

There have been attempts to develop automated systems for semen analysis since the 1980s, ever since it was possible to capture and process digital videos \cite{mortimer2015future}. The expectation from CASA systems was to have more objective and reproducible test results but there has been challenges in doing so \cite{mortimer2015future}. One of the reasons for unreliable results has been other particles and cells floating in the seminal plasma that cross the spermatozoa trajectory. The results improved when spermatozoa were removed from the seminal plasma and suspended in another medium, but such procedures require extensive training and preparation procedures. Despite the long history of CASA, it is due to these reasons that the use of CASA systems is not recommended by the WHO for clinical use \cite{mortimer2015future}\cite{world2010department}.

Urbano \textit{et al.} \cite{7748508} presented a fully automated multi-spermatozoa tracking system, which can track hundreds of spermatozoa simultaneously. It is also capable of measuring motility parameters with minimum human intervention. The method works by applying a modified version of the jpdaf algorithm to microscopic semen recordings, allowing them to track individual spermatozoon at proximities and during head collisions (a common issue with existing CASA instruments). The main contribution made by Urbano \textit{et al}. is the modified jpdaf algorithm for tracking individual spermatozoon, but they only tested the algorithm on two samples, so the generalizability of their approach cannot be determined. 

There were also other approaches like Mohammadi et al.\cite{mohammadi2020sperm} modified the CSR-DCF algorithm to perform individual spermatozoon tracking and predict quality parameters. Thambawita et al.\cite{thambawita2019extracting} used autoencoders to extract temporal features from the video sample into a spatial domain and used those features to predict motility. In another paper Thambawita et al.\cite{thambawita2019stacked} used stacked dense optical flows to predict sperm motility and morphology.

Hicks et al.\cite{hicks2019machine} tested various machine learning and deep learning methods on the VISEM \cite{Haugen:2019:3304109.3325814} dataset to predict spermatozoa motility. They used simple machine learning methods like linear and logistic regression, they also used deep learning methods like convolutional neural networks. Their deep learning methods gave consistent results and generalized well. Their results indicate that deep learning based methods would be the most feasible option in building a consistent CASA system. One issue with their method was that they used two-dimensional CNNs to process the videos frame by frame. This didn’t allow their model to use time as a parameter to learn better by connecting subsequent features. They also used 30 frames per video which constitutes to less than one second of data in a two to five minutes video. We improved upon the work of Hicks et al.\cite{hicks2019machine} with three-dimensional CNNs since spermatozoa motility depends primarily upon speed, incorporating time would help to capture the interrelated features much better between each consecutive frame.

\section{\normalfont \textsc{Proposed Approach}}
We took three variants of 3D Residual networks as follows.

\begin{figure}[htp]
    \centering
    \subfloat[3D ResNet18]{%
        \includegraphics[height=0.5\textwidth]{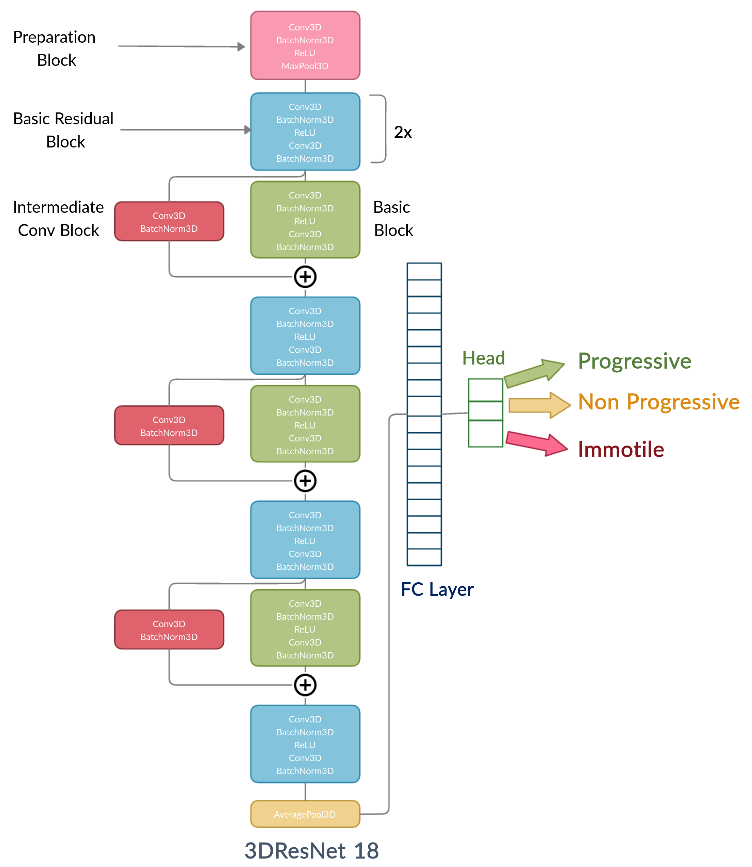}%
        \label{ResNet18}%
        }%
     \hfill%
    \subfloat[3D ResNet18 + Tabular Data]{%
        \includegraphics[height=0.5\textwidth]{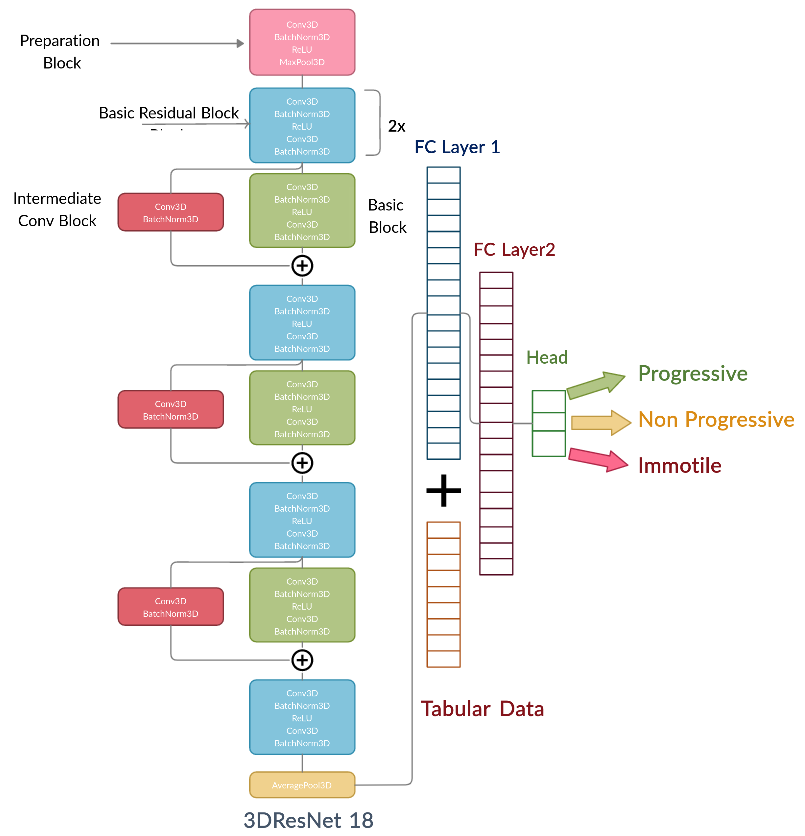}%
        \label{ResNet18 + Tabular Data}%
        }%
	\hfill%
    \subfloat[3D ResNet34 + Tabular Data]{%
        \includegraphics[height=0.5\textwidth]{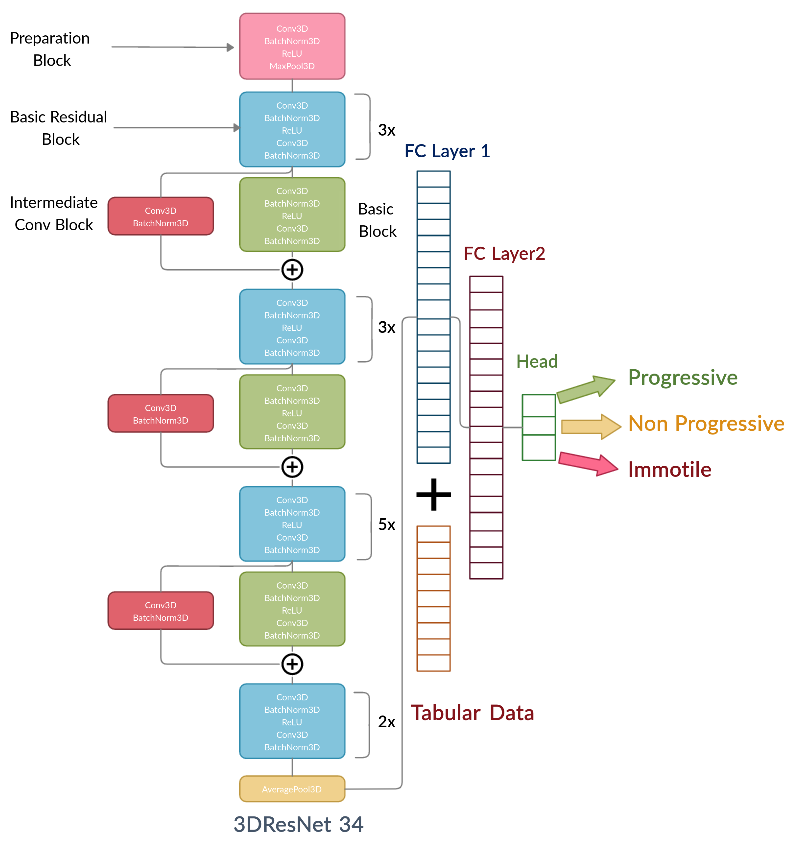}%
        \label{ResNet34 + Tabular Data}%
        }%
   \caption{Proposed Model Architectures}
\end{figure}

\subsection{\normalfont \textsc{3D ResNet18 }}
This model consists of 18 Convolutional layers, grouped together into blocks starting with the Preparation Block which consists of a 3D Convolutional layer of kernel size 7x7x7 with stride 1x2x2 and padding 3x3x3 that increases the channels from 1 to 64, a 3D Batch Normalisation layer applied across those 64 channels and passed through the ReLU activation function on which we apply a 3D Max Pooling layer with kernel of size 3x3x3. 

Next is Layer 1, which consists of two Basic Residual Blocks containing a 3D Convolutional layer of kernel size 3x3x3 with stride 1x1x1 and padding 1x1x1 in which the number of channels are maintained as 64. These blocks have internal residual connections. Layer 2 consists of two Basic Blocks and an Intermediate Conv Block. The output of Layer 1 is fed into one Basic Block and an Intermediate Conv Block (number of channels are increased from 64 to 128), the output of these two blocks are added and passed to the second Basic Block. The number of channels in this layer is maintained as 128. Layer 3 and Layer 4 consists of the same blocks and connections as Layer 2 except that the number of channels is maintained as 256 in Layer 3 and 512 in Layer 4. The output of Layer 4 is passed through a 3D Average Pool layer of kernel size 4x15x20 with stride 1 and padding 0. 

The output is passed to a 512 unit dense fully connected linear layer and our logit vector representing the three classes (progressively motile, non-progressively motile and immotile) is obtained.

\subsection{\normalfont \textsc{3D ResNet18 + Tabular Data}}
The structure of the 3D ResNet18 till the 3D Average Pooling layer remains the same (\ref{ResNet18}). The output of the Average pooling layer is connected to a dense fully connected layer of 512 units to which we augment the data of 19 most significant features (Seminal plasma anti-Müllerian hormone, Serum total testosterone, Serum oestradiol, Serum sex hormone-binding globulin, Serum follicle-stimulating hormone, Serum Luteinizing hormone, Serum inhibin B, Serum anti-Müllerian hormone, Abstinence time, Body mass index, Age, Sperm concentration, Ejaculate volume, Sperm vitality, Normal spermatozoa, Head defects, Midpiece and neck defects, Tail defects, Teratozoospermia index) of a given participant. We selected these features exclusively since they showed most correlation with motility prediction of a spermatozoon. Then we pass this 531(512 + 19) unit fully connected dense layer outputs to an 84 unit dense fully connected layer which outputs a logit vector that represents our three classes.

\subsection{\normalfont \textsc{3D ResNet34 + Tabular Data}}
We improve upon the previous model by taking a 34 Convolutional layers deep ResNet grouped together into blocks starting with the preparation block (\ref{ResNet18}). Next is Layer 1 in which then we take three Basic Residual Blocks (\ref{ResNet18}) with internal residual connections after which comes Layer 2 containing one Intermediate Conv Block, one Basic Block and three Basic Residual Blocks. The output of Layer 1 is fed into one Basic Block and the Intermediate Conv Block (\ref{ResNet18}), the outputs of these two blocks are added and passed into the Basic Residual Blocks. Layer 3 and Layer 4 contains one Intermediate Conv Block and one Basic Block with the same connections as Layer 2 along with five Basic Residual Blocks in Layer 3 and two Basic Residual Blocks in Layer 4. The output of Layer 4 is passed to a 3D Average Pool layer of kernel size 4x15x20 with stride 1 and padding 0. The output is connected to a dense fully connected linear layer of 512 units to which we augment 19 significant features from the tabular data (\ref{ResNet18 + Tabular Data}), and pass it through another fully connected layer to obtain our logit vector (\ref{ResNet18 + Tabular Data}).

\section{\normalfont \textsc{Implementation and Results}}

\subsection{\normalfont \textsc{Dataset}}

We use the VISEM dataset \cite{Haugen:2019:3304109.3325814}, a multi-modal dataset containing anonymized data from 85 different male participants which includes videos, biological analysis data, and participant data. The total size of the videos are over 35 gigabytes, the length of each video being between 2 and 7 minutes. Apart from the videos, there are six files containing tabular participant data. The details of the six files are given below - 
\begin{itemize}
	\item \textbf{semen\_analysis\_data}: The results of standard semen analysis such as sperm concentration and head defects.
	\item \textbf{fatty\_acids\_spermatozoa}: The levels of several fatty acids in the spermatozoa.
	\item \textbf{fatty\_acids\_serum}: The serum levels of the fatty acids of the phospholipids.
	\item \textbf{sex\_hormones}: The serum levels of sex hormones measured in blood such as serum inhibin B and serum oestradiol.
	\item \textbf{participant\_related\_data}: General information about the participants such as age, abstinence time and Body Mass Index.
	\item \textbf{videos}: Participant ID and video data mapping.
\end{itemize}

\subsection{\normalfont \textsc{Experimental Setup}}
We created an input pipeline where a participant’s video data is taken and the first 50 frames were extracted from it, an improvement from the 30 frames per video datapoint used by Hicks[ref]. We compiled these frames into a video tensor of 3 channels RGB. Then we iterated over these individual frames and converted each of them to channel-wise greyscale for colour isn’t an important feature in spermatozoa motility prediction and reducing the size of the video data by one third makes it an optimized solution for less powerful hardware. Then we stacked these greyscale frames together and fed it to our models. Since the dataset was highly imbalanced with a class distribution as follows - 

\begin{enumerate}
	\item \textbf{Progressively Motile} : 52
	\item \textbf{Non-Progressively Motile} : 9
	\item \textbf{Immotile} : 24
\end{enumerate}

We used weighted classes to prevent overfitting on one class. While training to avoid exploding gradients, we used grad\_clip\footnote{https://pytorch.org/docs/stable/generated/torch.nn.utils.clip\_grad\_value\_.html} of 0.1 and weight\_decay of 0.0001 along with a one fit cycle lr\_scheduler in PyTorch with a maximum learning rate in the range 0.1 to 0.001. We used Adam \cite{kingma2014adam} as the optimizer. The output logits from the models were passed through the CrossEntropyLoss function\footnote{https://pytorch.org/docs/stable/generated/torch.nn.CrossEntropyLoss.html} 
\begin{equation}
	\operatorname{loss}(x, \text { class })=\text { weight }[\text { class }]\left(-x[\text { class }]+\log \left(\sum_{j} \exp (x[j])\right)\right)
\end{equation}
to calculate the gradients and predict if the datapoint is progressively motile, non progressively motile and immotile.

All the experiments were done using PyTorch\footnote{https://pytorch.org/} and Pandas\footnote{https://pandas.pydata.org/docs/}. The code to reproduce the results can be found in our Gitlab repository\footnote{https://gitlab.com/innerve/sperm-motility-analysis/}.

\subsection{\normalfont \textsc{Results}}
We did a random seeded train, validation, test split on the dataset as follows -
\begin{itemize}
	\item Train set : 63 samples.
	\item Validation set : 8 samples.
	\item Test set : 9 samples.
\end{itemize}

\begin{figure}[htp]
    \centering
    \subfloat[3D ResNet18]{%
        \includegraphics[width=0.33\textwidth]{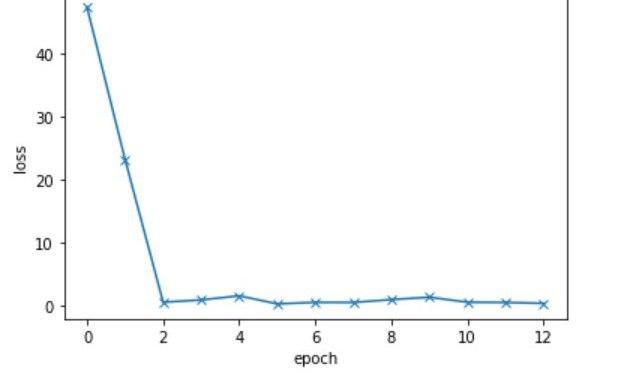}%
        \label{3D ResNet18 loss}%
        }%
     \hfill%
    \subfloat[3D ResNet18 + Tabular Data]{%
        \includegraphics[width=0.33\textwidth]{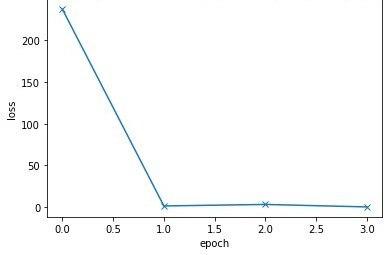}%
        \label{3D ResNet18 + Tabular Data loss}%
        }%
	\hfill%
    \subfloat[3D ResNet34 + Tabular Data]{%
        \includegraphics[width=0.33\textwidth]{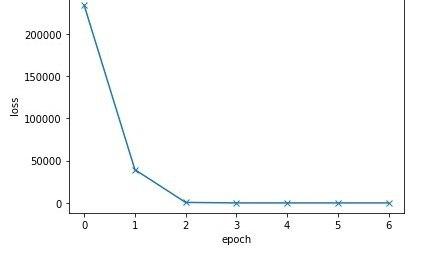}%
        \label{3D ResNet34 + Tabular Data loss}%
        }%
   \caption{Loss Graphs}
	\label{loss Graphs}
\end{figure}

The first ResNet18 model reached 100\% accuracy on both validation and test
sets. The second ResNet18 + Tabular Data model correctly predicted with
an accuracy of 100\% on the validation set and an average accuracy 88.89%
on the test set. The third ResNet34 + Tabular Data model correctly predicted
with an accuracy 87.5\% on the validation set and 77.78\% on the test set. We
used early stopping as a criteria for all models to avoid overfitting where we
stopped training the first and second model after the 12 th epoch and the third
model after the 6 th epoch, where ultimately the loss curve flattens out as shown
in figures 6, 7 and 8.

VISEM \cite{Haugen:2019:3304109.3325814} provided a baseline Mean Absolute Error metric for all regression models trained on this dataset. But since our experimental setup used classification for predicting spermatozoa motility, making it the first classification approach taken with this dataset, we cannot directly compare the results with the baseline score given in the dataset. The novel accuracy we established paves a way towards future classification tasks and improvements.

\section{\normalfont \textsc{Conclusion}}

We proposed a novel method to classify spermatozoa motility using spatial
temporal 3D Convolutional Neural Networks from significantly less video
data (50 frames per video datapoint). Our experiments concluded that the addition of tabular data decreased the accuracy of the model (\ref{3D ResNet18 + Tabular Data loss}), which isn't necessarily an architecture limitation as the accuracy decreased further when we used a heavier 3D ResNet34 (\ref{3D ResNet34 + Tabular Data loss}). We found that these features were redundant for the training and classification of spermatozoa motility as they cause data leakage. The proposed models can be applied to automating the semen analysis process in the future which currently uses images instead of video and can even extend to predicting other quality parameters.

\bibliographystyle{unsrt}  
\bibliography{references}

\end{document}